\documentclass[%
reprint,
amsmath,amssymb,
prst-ab,
]{revtex4-1}
\usepackage{amsmath}
\usepackage{cases}
\usepackage{graphicx}
\usepackage{dcolumn}
\usepackage{bm}
\usepackage{textcomp}
\usepackage{color}

\newcommand{\re}[1]{{\color{black}#1}}

\begin{document}

\preprint{APS/123-QED}

\title{Ion Acceleration \re{via ``Nonlinear Vacuum Heating" by
the Laser Pulse Obliquely Incident on a Thin Foil Target} 
}

\author{A. Yogo$^{1,2}$}
\author{S. V. Bulanov$^{1}$}
\author{M. Mori$^{1}$}
\author{K. Ogura$^{1}$}
\author{T. Zh. Esirkepov$^{1}$}
\author{A. S. Pirozhkov$^{1}$}
\author{M. Kanasaki$^{1}$}
\author{H. Sakaki$^{1}$}
\author{Y. Fukuda$^{1}$}
\author{P. R. Bolton$^{1}$}
\author{H. Nishimura$^{2}$} 
\author{K. Kondo$^{1}$} 
\affiliation{$^{1}$Kansai Photon Science Institute, Japan Atomic Energy Agency, 
Kizugawa 619-0215, Kyoto, Japan}
\affiliation{$^{2}$Institute of Laser Engineering, Osaka University, Suita 565-0871, Osaka, Japan}

\date{\today}

\begin{abstract}
\re{Dependence of the energy of ions accelerated during interaction of the laser pulse obliquelly 
incident on the thin foil target on the laser polarization is studied experimentally and theoretically.
We found that the ion energy being maximal for the p-polarization gradually decreases when the 
pulse becomes s-polarized. The experimentally found dependences of the ion energy are explained by 
invoking the anomalous electron heating which results in high electrostatic potential formation at the 
target surface.
}
Anomalous heating of electrons beyond the energy of quiver motion in the laser field \re{is described 
within the framework of theoretical model of driven oscillator with a step-like nonlinearity.}
\re{ We have demonstrated that the electron anomalous heating can be realized in two regimes:}
nonlinear resonance and stochastic heating, depending on the extent of stochasticity.
\re{We have found the accelerated ion energy scaling determined by the laser}
intensity, pulse duration, polarization angle and incident angle.
\end{abstract}

\maketitle


\section{Introduction}

Discovery of laser-driven ion acceleration \cite{review} has opened a broad field of potential 
applications involving fast ignition \cite{ROTH2001, IonsGUS}
and hadron therapy \cite{therapy1, therapy2, HT-UFN}.
Most intensively investigated has been the mechanism of ion acceleration with charge 
separation (CS) fields generated on the thin-foil target surfaces, including target 
normal sheath acceleration (TNSA) \cite{review}.
Nowadays, increasing attention is also paid for different mechanisms, 
however, the TNSA mechanism still remains crucial to understanding physics of ion acceleration at the moderate 
laser intensity regimes.

Generation of ion accelerating CS field is governed by the absorption mechanism of laser energy into electrons. 
In the case of interactions between obliquely incident laser and overdense plasma, 
one of the most predominant absorbed mechanism is the \textit{Brunel} effect \cite{brunel}.
In this model, electrons on the boundary are directly accelerated by the electric field of the laser, 
when the electrons are accelerated to the normal direction of the surface with a momentum 
$p_{max} = 2m_{e}ca_{0}$ at maximum, where $c$ is the speed of light in vacuum, 
$m_{e}$ and $e$ are the electron mass and charge; $a_{0} = e E_{l}/m_{e}\omega c$ 
is a dimensionless amplitude of the laser \re{radiation}, 
with $E_{l}$ and $\omega$ being the laser electric field amplitude and frequency.
\re{It can be expressed in terms of the laser intensity $I$ and wavelength $\lambda=2\pi c/\omega$ as 
$a_{0} = 0.85 \sqrt{I (10^{18} \mbox{ W/cm}^2) \lambda^2 (\mu \mbox{m}^2)}$. }

To explain the acceleration beyond $p_{max}$, nonlinear effect should be \re{implemented} into the analysis.
D'yachenko and Imshennik \cite{Dyachenko} numerically revealed that electrons quivered by the laser field
are kicked-out by the potential boundary of the plasma and anomalously gain momentum depending on time 
(see also \cite{RECIRC, LMM-2015}).
Paradkar \textit{et al.} \cite{Paradkar} and Krasheninnikov \cite{Krasheninnikov} theoretically showed 
stochastic heating of electrons by assuming a model \re{dependence on coordinate of the} 
electric field corresponding 
to the field in the vicinity of a thin-foil targets.
Using particle-in-cell \re{simulations}, Taguchi \textit{et al.} \cite{Taguchi} found that 
the heating of electrons 
transiting through a small size target can be higher than that by the electron 
quivering and called this phenomenon as ``nonlinear resonance absorption." 
Robinson \textit{et al.} \cite{robinson} proposed non-wake-field acceleration 
beyond the momentum limit is gas plasma. 

In this manuscript, we develop the mechanism of ion acceleration involving nonlinear 
effect on the electron energization in order to describe the accelerated proton 
energy during the laser interaction with the thin foil target.
By considering a model of driven oscillator with a step-like nonlinearity, 
the growth of chaotic behaviour of electrons is revealed.
Here, we \re{call}  the mechanism  \textit{the Nonlinear Vacuum Heating}.
As a result, the experimental results on the proton energy and its dependency 
on the laser polarization are successfully predicted.

\section{Experiments}

The experiment was performed using JLITE-X laser at JAEA, 
which delivers ultrashort ($\tau = 40$ fs, FWHM) linearly-polarized 
pulses having a central wavelength of $\lambda = 800$ nm.
The laser contrast has been raised to $10^9$ by the benefit 
of Insertable Pulse Cleaning Module \cite{IPCM}.
The beam is focused to a spot of 8 $\mu$m (FWHM) 
in diameter using an $f = 160$ mm off-axis parabolic mirror (OAP) 
with an incidence angle of 45$^{\circ}$ on silicon-nitride foils.
The polarization angle is varied from \textit{p} ($\theta = 0^{\circ}$) to \textit{s} ($\theta = 90^{\circ}$) 
by using a half-waveplate located at the entrance of the OAP.
The areal peak intensity obtained is $I = 3.0 \times 10^{18}$ W/cm$^{2}$, 
which corresponds to the dimensionless amplitude $a_{0} = 1.2$.
The pedestal intensity level in sub-ns range is around $10^9$ W/cm$^{2}$ 
that is adequate to suppress the preplasma formation on the ultrathin foil surface.
Accelerated protons are measured by Time-of-Flight (TOF) online detector \cite{TOF} 
located at the normal direction of the target rear surface.
\re{It is confirmed using ion-track (CR-39) detection that
the protons ($> 1.2$ MeV) has a 15$^{\circ}$ angular divergence 
distributed along the normal of the target surface \cite{IPCM}.}

\begin{figure}[]
  \begin{center}
    \includegraphics[keepaspectratio=true,height=65mm]{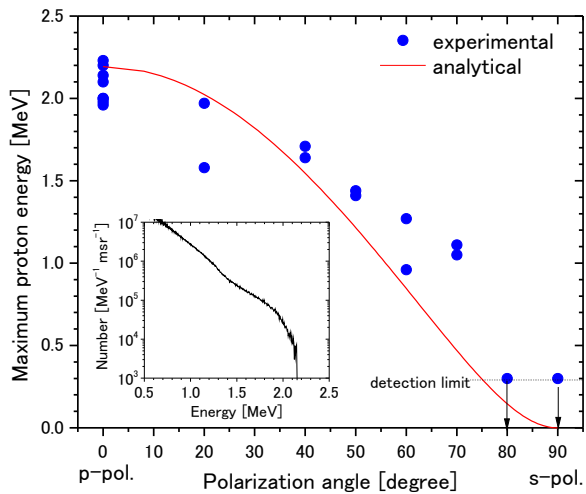}
  \end{center}
  \caption{Maximum proton energies (circles) observed for 500 nm 
  thick targets as a function of the polarization angle $\theta$. 
\re{  When the laser polarization is close to \textit{s}-polarization 
($\theta \ge 80^{\circ}$), 
  the proton energy becomes below than the detection limit. }
  Analytically predicted proton energy ${\cal E}_{p}$ is shown with a solid line. 
  The inset shows a typical energy spectrum for $\theta = 0$.}
  \label{Fig1}
\end{figure}

\re{In the experiment we analyze the dependence of the energy of accelerated ions 
on the thin foil target on the laser polarization.}
Figure \ref{Fig1} shows the maximum energy of protons (circles) observed for 500-nm-thick 
targets as a function of the polarization angle $\theta$.
Here, the proton energy is drastically decreasing as the polarization 
is changed from \textit{p} to \textit{s} mode.
Similar dependency was also observed \cite{Ceccotti} for a thicker (13 $\mu$m) 
target under high-contrast laser conditions.
On the other hand, there is a clear difference from the 
case of low-contrast laser \re{when} the proton energy 
for \textit{s}-polarization was only by 30\% lower  \cite{fukumi} 
than that for \textit{p}-polarization. 
\re{As we see}, the obvious dependency on the polarization angle 
is characteristic of the interaction between steep-gradient 
plasma surfaces and obliquely-incident laser pulses.
To understand the physics underlying the phenomenon, 
we discuss the  motion of \re{electrons driven by a strong EM wave 
in the vicinity of }the plasma-vacuum interface.

\section{Oblique incidence of the laser radiation on a thin foil target}

\subsection{Electron quiver motion}

\begin{figure}[]
  \begin{center}
    \includegraphics[keepaspectratio=true,height=25mm]{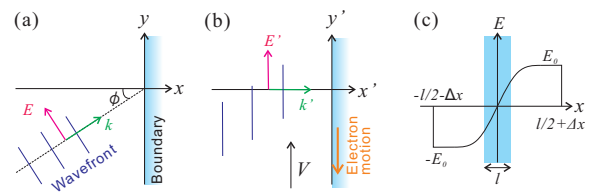}
  \end{center}
  \caption{\re{Electromagnetic wave obliquelly incident 
  on the thin foil target. (a) The wave in the laboratory frame of reference $K$.
  (b) The wave in the boosted frame of reference $K'$. 
  For the \textit{p}-polarization wave, the electric field vector is in the $x$-$y$ plane. 
  In the boosted frame of reference, the plasma electrons move 
  along the target with a velocity $-V = -c \sin{\phi}$ in the $y'$ direction. 
  (c) A charge separation electric field on the $x$ axis.
  }
  }
  \label{Fig2}
\end{figure}

Let us consider a plane electromagnetic (EM) wave interacting with 
the thin foil located in the plane $x=0$. \re{The laser the angle of incidence equals 
$\phi$ (see Fig. \ref{Fig2} (a))}. 
The EM wave is linearly polarized and its electric field $\textbf{E}$ has a polarization 
angle $\theta$ with respect to the $x$-$y$ plane: $\theta=0$ for \textit{p}-polarization 
and $\theta=90^{\circ}$ for \textit{s}-polarization.
In order to simplify the consideration, we perform a Lorentz transformation from 
the laboratory frame $K$ to the frame of reference $K'$ moving with the velocity $V = c \sin{\phi}$ 
along the target surface (y direction).
In the $K'$ frame, the EM wave is incident normally onto the surface \cite{Bourdier,Vshivkov}, 
and the gamma factor associated with the velocity $V$ is given by 
\begin{equation}
\gamma_{M} = \frac{1}{\sqrt{1-V^2/c^2}} = \frac{1}{\cos{\phi}}.
\end{equation}
Fourier component for a monochromatic plane wave in vacuum \cite{Landau} are given in $K'$ frame 
\re{by equations}
\begin{equation}
\nabla \times \textbf{E}' = i \frac{\omega'}{c} \textbf{B}'
\label{eq2} 
\end{equation}
\begin{equation}
\nabla \times \textbf{B}' = -i \frac{\omega'}{c} \textbf{E}',
\label{eq3}
\end{equation}
where the wave frequency \re{equals} $\omega' = \omega/\gamma_{M} = \omega \cos{\phi}$.
Using $\nabla \cdot \textbf{E}' =0$, the Eqs. (\ref{eq2}) and (\ref{eq3}) \re{can be written as follows}
\begin{equation}
\nabla ^{2} \textbf{E}' + \frac{\omega'^2}{c^2} \textbf{E}' = 0.
\label{eq4}
\end{equation}
Using Eqs. (\ref{eq2}--\ref{eq4}) we obtain the general solution of the Maxwell equations, 
\begin{equation}
\textbf{E}' = \textbf{E}_{0} e^{i \omega'(t'-x'/c)} + \textbf{E}_{r} e^{i \omega' (t'+x'/c)}
\end{equation}
with the electric field amplitudes of incident $\textbf{E}_{0}$ and 
reflected $\textbf{E}_{r}$ waves in the $K'$ frame. 

When the EM wave interacts with the plasma having a steep gradient satisfying the condition $r_{E}/L \gg 1$,
we can use \re{the Leontovich}  boundary conditions \cite{LLECM} for the tangential components of the
electric and magnetic fields $\textbf{E}_t$ and $\textbf{B}_t$:
\begin{equation}
\textbf{E}_t = \zeta \textbf{B}_t \times \textbf{n},
\end{equation}
where $\textbf{n}$ is the unit vector along the inward normal 
to the plasma surface and $\zeta$ is the surface impedance operator.
\re{Here $r_E$ is the electron quiver radius and $L$ is the scale-length 
of the plasma inhomogeneity}.
Assuming that the plasma surface can be treated as an ideal conductor ($\zeta=0$), 
we obtain the boundary condition
\begin{equation}
\textbf{E}'_{x'=0} = \textbf{E}_t = 0.
\end{equation}
Then, the electric field is written as
\begin{equation}
\textbf{E}' = \textbf{E}_{0} [e^{i \omega'(t'-x'/c)} - e^{i \omega' (t'+x'/c)}].
\label{eq8}
\end{equation}
The magnetic field is obtained using Eqs.(\ref{eq2}) and (\ref{eq8}) as follows:
\begin{equation}
\textbf{B}' = \textbf{E}_{0} [e^{i \omega'(t'-x'/c)} + e^{i \omega' (t'+x'/c)}].
\end{equation}
Therefore, the real part of the magnetic field at the surface ($x'=0$) is given by
\begin{equation}
\textbf{B}'_{x'=0} = (0, \ -2 E_0 \sin \theta \cos{\omega' t'}, \ 2 E_0 \cos{\theta} \cos{\omega' t'}).
\end{equation}

In the boosted frame of reference, $K'$, the acceleration of electrons in the $x'$ direction 
is driven by the \re{Lorentz force equal to $-({e}/{c}) \textbf{v} \times \textbf{B}$,
where the magnetic field is taken at the $x=0$ plane, $\bf{B}=\textbf{B}'_{x'=0}$, 
and the velocity of the electrons is equal to $\textbf{v} =(0, -V, 0)$.}
As the result the equation of motion for electron \re{can be written as}
\begin{equation}
\frac{d\textbf{p}'}{dt'} = -\frac{e}{c} [\textbf{v} \times \textbf{B}'_{x'=0}],
\end{equation}
\re{It yields for the $x'$-component of the electron momentum}
\begin{equation}
\frac{dp_{x'}}{dt'} = 2 m_{e} V a_{0} \omega'\cos{\theta} \cos{\omega' t'}.
\end{equation}
Hence,  the $x'$-component of the electron momentum \re{depends on time as}
\begin{equation}
p_{x'} =  2 m_{e} V a_{0} \cos{\theta} \sin{\omega' t'},\label{s}
\end{equation}
where 
is the dimensionless amplitude of the laser and $V = c \sin{\phi}$.
\re{We see that the electron momentum $p_{x'}$ is below $2 m_{e} c a_{0}$, 
which corresponds 
to the momentum limit for electrons accelerated by obliquely incident laser via the Brunel effect \cite{brunel}.}

\subsection{Nonlinear mechanism of anomalous electron heating}

\re{Within the framework of the Nonlinear Vacuum Heating mechanism, 
the electrons quivered 
by the laser field are \textit{kicked-out} by the steep-like electric field 
at the plasma boundary. As the result, the EM field energy is irreversibly converted into 
the electron kinetic energy, thus enhancing the electron heating.}
To \re{describe} the nonlinear motion of electrons, we assume \re{that
a step-like CS electric field is formed near the ion layer having the positive electric charge. }
The equations of electron motion can be written in the form
\begin{equation}
\dot p+\varepsilon\, {\rm sign}{(x)}=a_{0} \cos t,
\label{eq:dotp}
 \end{equation}
\begin{equation}
\dot x=\frac{p}{\left(1+p^2\right)^{1/2}}.
\label{eq:dotx}
 \end{equation}
\rm{Here the sign function is equal to} ${\rm sign}{(x)}=1$ for $x>0$ and ${\rm sign}{(x)}=-1$ for $x<0$, 
and a dot denotes a differentiation with respect to time, $t$.
\rm{The electron momentum is normalized on $m_e c$, the time is measured in units $\omega^{-1}$, 
and the coordinate is normalized on $c/\omega$.}

 The parameter 
\begin{equation}
\varepsilon_p=2\pi n_0 e^{2} l/m_e \omega c,
\label{eq:eps_p}
 \end{equation}
 introduced in Ref. \cite{Vshivkov}, is proportional to the charge separation electric field, 
$E_{0}=2\pi en_0 l$ (see Fig. \ref{Fig2} (c)), normalized on $m_e\omega c/e$, for a thin foil with 
the electric charge surface density equal to  $en_0l$. 

\re{The formulated model of a driven oscillator with a step-like 
nonlinearity (see also \cite{Bulanov2015}) helps to determine and elucidate the conditions 
when the stochastic vacuum heating takes place. The  3-dimensional 
model used bears the key features of the problem 
addressed in Refs. \cite{Paradkar, Krasheninnikov, ZMS-2002, BPL-2007}, but in a significantly  
simpler mathematical approximation, 
because a full description of the 
charged particle interaction with electromagnetic waves implies 
a consideration of a 7-dimensional dynamical system.}

While the electron trajectory crosses the $x=0$ plane, 
the sign of the static electric field changes abruptly. 
This can be considered as a ``collision'', during which the oscillating 
driver electric field produces the work changing the electron energy. 
Depending on the phase, when the ``collision'' happens, the work can be either positive or negative, 
resulting in the electron energy increasing or decreasing. 

The enhancement of the energy transfer from the driver electric field to the electron 
is expected to occur when a resonant condition \re{(for details see Ref \cite{Bulanov2015})} takes place:
\begin{equation}
a_{0} = \pi \varepsilon/2,
 \end{equation}
which can be equivalent to the relationship of $E_{0} \approx E_{l}.$
The expression for the momentum maximum is found in Ref. \cite{Bulanov2015} 
for two different nonlinear regimes: \textit{nonlinear resonance} and \textit{stochastic heating}. 

In nonlinear resonance regime, electrons gain a momentum $\Delta p \approx a_{0}$ 
during the passage of the plane $x=0$.
Then, the momentum grows  proportionally to the square root of time, written in
\begin{equation}
p_m \approx  (a_0 \varepsilon t/2)^{1/2}=a_0 (t/\pi)^{1/2}
\label{eq:p-treg}
 \end{equation}
for the regime of nonlinear resonance.
In the case of stochastic heating, the momentum dependence on time 
has a character of diffusion with the diffusion coefficient 
\begin{equation}
D_{pp}=\frac{2 \Delta p \Delta p}{T}=\frac{a_0^2 \varepsilon}{2 p_m}.
\label{eq:Dpp}
 \end{equation}
This yields for the time dependence of the maximum electron momentum
\begin{equation}
p_m \approx \left( a_0^2 \varepsilon t/2\right)^{1/3}=a_0 \left( t/\pi \right)^{1/3},
\label{eq:p-tstoch}
 \end{equation}
 noticed in \cite{Paradkar}.

To analyze in more details the \re{ properties of nonlinear oscillations} we have 
integrated Eqs. (\ref{eq:dotp}, \ref{eq:dotx}) numerically.
For regularization of the singularity in the left hand side of Eq. (\ref{eq:dotp}) 
we use instead the sign function ${\rm sign}(x)$ the function ${\rm Tanh}(x/l)$ 
with the ion layer width $l$ (see Fig. \ref{Fig2} (c)) 
substantially small compared to the particle displacement: $l \le \Delta x.$
\begin{figure}[]
  \begin{center}
    \includegraphics[keepaspectratio=true,height=170mm]{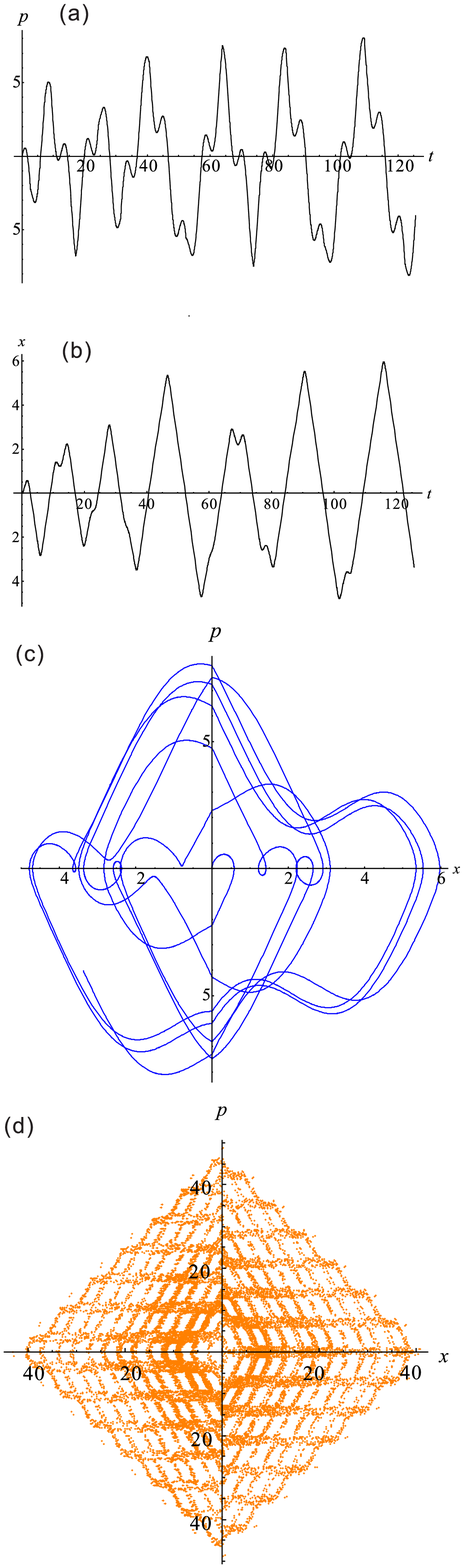}
  \end{center}
  \caption{Driven ocillations obtained for $a_0 = 2$ and $\varepsilon_p = 1.35$. 
  Normarized momentum (a) and coordinate (b) of electrons as vs time. 
  (c) The phase plane $(x,p)$. 
  (d) The Poincar\'{e} section showing the particle 
  positions on the phase plane at the discrete time with the time step 
equal to the period of the driving force, $2 \pi$.}
\label{Fig3}
\end{figure}
Figures \ref{Fig3} (a,b) show the result of numerical calculation of the momentum 
and the coordinate of electrons as functions of time obtained for the condition close to our experiment.
\re{We can see clearly that the oscillations are nonlinear with the momentum abruptly 
droping down and rising up again. 
In a time duration of the laser pulse, 40 fs ($t = 95$ in Fig. \ref{Fig3}), the momentum 
reaches $8 m_{e}c$, which is a few times higher than the quiver momentum corresponding to the Brunel effect}.
The oscillation amplitude is limited both in the momentum and coordinate directions, 
shown in Fig. \ref{Fig3}(c).
The trajectory tightly fills the domain in the phase plane demonstrating ergodic property 
of the system under consideration.
In the Fig. \ref{Fig3} (d) we plot a Poincar\'{e} section having a form of finite thickness web, 
which is broadened due to stochastic nature of the particle motion.
The property of the particle to migrate on the phase plane along the web is typical 
for the minimal chaos regimes \cite{Chernikov}.
In other words, chaotic behavior has begun to emerge, but not evolves completely under these conditions.

\subsection{Ion acceleration}

Nonlinear Vacuum Heating conjecture is equivalent to the assumption that 
the energy conversion from the EM field to electron energy depends on time according 
to the nonlinear mechanism introduced above.
The average square of $x'$-component of the electron momentum $p_{nvh,x'}$ scales 
with \re{the laser pulse  duration which is proportional to the number 
of wave periods $N$} for the two different 
regimes of nonlinear vacuum heating as:
\begin{numcases}
{\overline{p^2}_{nvh,x'} \propto} N \overline{p^2_{x'}} & \text{(nonlinear resonance)}\label{pnlr}\\
 N^{2/3} \overline{p^2_{x'}} & \text{(stochastic heating)}
\label{pstc}
\end{numcases}

To estimate the typical energy gain of fast protons, we  use 
the step-like CS field equal to $ E_{0} \tanh(x/l)$ in the interval 
\begin{equation}
-l/2-\Delta x \le x \le l/2+\Delta x,
 \end{equation}
where $\Delta x$ is the electron displacement and $l$ is the thickness of positively charged 
layer of the foil target  (Fig. \ref{Fig2}(c)).
When the target thickness is thin enough to satisfy the condition $l \simeq \Delta x$,
the electric field $E_{0}$ can be expressed as
\begin{equation}
E_{0} = 2 \pi n_{0} e l = 2 \pi n_{0} e \Delta x\label{E0}
\end{equation}
with the ion charge density equal to $n_{0}$.
Assuming that protons can be treated as test particles initially located 
on the surface of the layer, the proton energy gain ${\cal E}_{p}$ can 
be estimated to be of the order of the electron kinetic energy, ${\cal E}_{e}$:
\begin{equation}
{\cal E}_{p} = e E_{0} \Delta x = 2 \pi n_{0} e^{2} \Delta x^{2} 
\simeq \frac{\overline{\Delta p^{2}}}{2m_{e}} = {\cal E}_{e}.
\end{equation}
In the boosted frame of reference $K'$, the kinetic energy can be found to be given by
\begin{eqnarray}
{\cal E}'_{p} &=&{\cal E}'_{e} = {\cal E}_{total}' - m_{e} c^2 \\
{\cal E}_{total}' &=& \frac{\omega'}{2 \pi}\oint \sqrt{m_{e}^2 c^4 + p_{V}^2c^2 + p_{nvh,x'}^2c^2} dt,
\end{eqnarray}
where 
\begin{equation}
p_{V} = \gamma_{M} m_{e} V = m_{e} c \, \tan{\phi}.
\end{equation}
From Eqs.(\ref{s}), (\ref{pnlr}) and (\ref{pstc}), follows that
\begin{equation}
p_{nvh,x'} = N^{j} p_{x'} = 2 N^{j} m_{e} a_{0} c \sin{\phi} \cos{\theta} \sin{\omega' t'}
\end{equation}
with the index $j$ equal to $j=1/2$ for the nonlinear resonance regime 
and equal to $j=1/3$ in the stochastic heating regime. 
The kinetic energy gain in the laboratory frame of reference, 
${\cal E}_{p}$ can be found by performing the inverse Lorentz transformation from the boosted to laboratory frame 
of reference,
\begin{equation}
{\cal E}_{p} = \gamma_{M}({\cal E}_{total}'-p_{V} V) - m_{e} c^2.
\end{equation}
This the equation for the proton energy ${\cal E}_{p}$. Its analytical solution yields
\begin{equation}
{\cal E}_{p}= \frac{m_{e} c^2}{\cos^2{\phi} } 
\left\{ \frac{1}{\pi}\left[ E({\cal A}) +E\left(\frac{{\cal A}}{1+{\cal A}}\right) 
\sqrt{1+{\cal A}} \right]-1 \right\},
\label{eq:EpK}
\end{equation}
where 
\begin{equation}
{\cal A}= a_{0}^2 N^{2j} \cos^2{\theta} \sin^2{2\phi}
\end{equation}
and 
\begin{equation}
E(z)=\int _{0}^{\pi/2} \sqrt{1-z^2 \sin^2{\Theta }} d\Theta
\end{equation}
 is the complete elliptic integral of the second kind \cite{GR}. 

It should be emphasized that the present model determines the maximum energy of protons  
as a function of the laser parameters: dimensionless amplitude $a_{0}$, 
number of wave period $N$, angles of polarization $\theta$ and incidence $\phi$.
In Fig. 1, the analytically value of proton energy ${\cal E}_{p}$ predicted with nonlinear 
resonance mechanism (see Eq.(\ref{eq:EpK}) with $j=1/2$) is shown as a function of $\theta$.
Here, we are using parameters employed in the experiment: $a_{0} = 1.2$, $\phi = 45^{\circ}$, 
and $N = 15$, which corresponds to the number of wave periods included 
in the FWHM duration ($\tau =40$ fs) of the incident laser.
On the other hand, the proton energy is also reproduced by stochastic heating 
(See Eq.(\ref{eq:EpK}) with $j=1/3$)  
using $(a_0, \phi, N) = (1.7, 45^{\circ},15)$, where the $a_0$ value 
is slightly higher than the experimental one.
Although it may be possible to assume that the enhancement of $a_0$ from 1.2 to 1.7 
is caused by a self-focusing of laser in preformed plasma, it is natural to consider 
that the electron energization in the present experimental condition is predominantly 
governed by nonlinear resonance mechanism, not by stochastic heating.
This aspect is consistent with the fact revealed by numerical analysis (Fig. 3) 
that the  process is in the intermediate stage between nonlinear resonance 
and stochastic heating and that chaotic behavior of energization has just begun to grow.
\re{Substantially, the chaotic regime requires a long time to develop 
as an asymptotic stage after the nonlinear resonance.}

\begin{figure}[]
  \begin{center}
    \includegraphics[keepaspectratio=true,height=50mm]{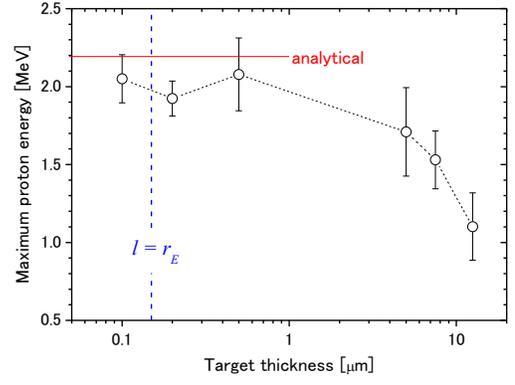}
  \end{center}
  \caption{Maximum proton energy observed for different values of the the target thickness: 
  0.1-0.5 $\mu$m thick silicon nitride and 5-12.5 $\mu$m thick polyimide. 
  Bars correspond to a double standard deviation ($2\sigma$) for each point. 
  The value of the proton energy ${\cal E}_{p}$  shown by a solid line 
  \re{is found from Eq. (\ref{eq:EpK})}.}
  \label{FIG-thick}
\end{figure}

As is shown in Eq.(\ref{E0}), the present model is valid when 
the target thickness satisfies $l \simeq \Delta  x$.
The displacement of electrons $\Delta x$ is typically assumed to be equal to the amplitude 
of electron oscillation $r_{E} = e E_{l}/m_{e}\omega^2$.
\re{Hence, the present model is applicable when the target thickness satisfies $l \simeq r_{E}.$
To demonstrate applicability of the criterion above, we observed the target 
thickness dependency of proton energy, shown in Fig. \ref{FIG-thick}.}

Fig. \ref{FIG-thick} shows \re{ the dependence 
of the proton energy on the target thickness 
observed with $(a_{0}, \theta,  \phi, N) = (1.2, 0, 45^{\circ}, 15)$.
The maximum proton energy is close 
to the value of analytical prediction (2.18 MeV) given by Eq. (\ref{eq:EpK})
when the target thickness is ranging around $r_{E} = a_{0} \lambda /2\pi = 0.15$ $\mu$m}.
For thicker targets ($l \ge 5 \mu$m), proton energy decreases with 
the increase of the target thickness $l$.
This can be attributed to the neutralization of ion density $n_{0}$ 
by the returning current, \re{ whose effect is subtantially stronger in the case of } thicker targets.

\subsection{Ion energy scaling}

\begin{figure}[]
  \begin{center}
    \includegraphics[keepaspectratio=true,height=125mm]{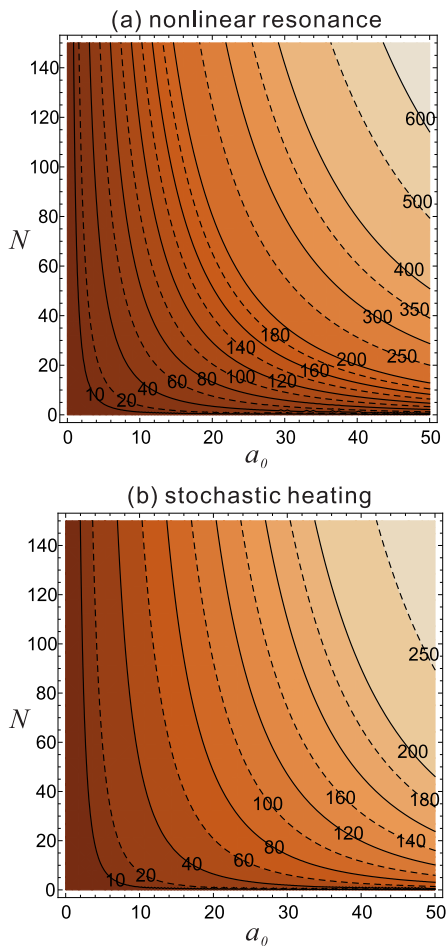}
  \end{center}
  \caption{\re{ Isocontours of the proton energy ${\cal E}_{p}$ measured in a unit 
  of MeV as in the plane of the laser amplitude $a_{0}$ and normalized duration $N$}
  in two different regimes of nonlinear vacuum heating: 
  (a)nonlinear resonance and (b) stochastic heating. 
  The laser incidence and polarization is fixed on $\phi=60^{\circ}$ 
  and $\theta = 0$.}
  \label{scale-fig}
\end{figure}
Figure \ref{scale-fig} shows the acquired proton energy according to Eq.(\ref{eq:EpK}), 
under the nonlinear resonance ($j=1/2$) (a) and stochastic heating ($j=1/3$) (b).
\re{Isocontours of the proton energy ${\cal E}_{p}$ measured in a unit 
  of MeV as in the plane of the laser amplitude $a_{0}$ and normalized duration $N$}.
In this figure, the proton energy is constant along a contour line $a_{0}^2 N^{2j}=$constant.
\re{It is interesting to note that the nonlinear 
resonance provides the energy scaling on $a_{0}^2 N$, 
which is proportional to the laser energy fluence.}

As is discussed above, the stochastic heating requires a substantially 
long time to develop being an asymptotic stage after the nonlinear resonance.
Therefore, when the laser pulse duration is relatively short, 
typically $ \tau \sim 50$ fs ($N \sim 20$ for $\lambda = 800$ nm) 
as used in our experiment, the scaling perhaps correspond to the nonlinear resonance regime.
\re{According to this hypothesis, the energy gain up to 200 MeV
(this is one of the milestones toward clinical applications of the laser accelerated ions \cite{HT-UFN}), 
can be achieved for the laser radiation with the amplitude $a_{0} = 42$ (see Fig. \ref{scale-fig}(a)). 
This laser amplitude corresponds to the intensity of $I = 4 \times 10^{21}$ W/cm$^{2}$ 
and  requires the target with optimal  thickness of the order of $l \simeq r_{E} = 5.4$ $\mu$m.}
\re{On the other hand side, 
a longer pulse duration, e. g.  when $\tau = 500$ fs for laser wavelength o
f $\lambda = 1.06 \, \mu$m, $N = 142$, the proton energy equal to 
200 MeV can be obtained with the laser amplitude equal to $a_{0} = 16$ 
corresponding to the laser intensity $I = 3 \times 10^{20}$ W/cm$^{2}$ and 
the target thickness of the order of $l \simeq 2.6\mu$m, provided 
 the stochastic heating mechanism does not occur.}
The laser parameters written above have been achieved in several laser facilities nowadays.
However, for a longer pulse duration 
(i.e. for $\tau = 500$ fs, $\lambda = 1.06 \, \mu$m $N = 142$) 
in the stochastic heating regime the 200-MeV proton acceleration 
requires the laser amplitude to be equal to $a_{0} = 34$ (see Fig. \ref{scale-fig}(b)) 
corresponding to the intensity $I = 1.6 \times 10^{21}$ W/cm$^{2}$.

\section{Conclusion}
In conclusion, 
\re{We have studied the ion acceleration during interaction 
of the short pulse laser obliquelly 
incident on the thin foil target. 
The dependence of the ion energy of on the laser polarization has been studied 
studied experimentally and theoretically.
We found that the ion energy being maximal for the p-polarization gradually decreases when the 
pulse becomes s-polarized. The experimentally found dependences of the ion energy are explained by 
invoking the anomalous electron heating, which results in efficient ion acceleration at the 
target surface.}

We showed that the proton energy can be substantially enhanced under the conditions of 
nonlinear resonance experienced by the electrons circulating around the thin foil target 
irradiated by the electromagnetic wave. 

\section*{Acknowledgments}
This work was founded by a Scientific Research (C) No. 25420911 
commissioned by MEXT and partially supported by NEXT Program of JSPS.
AY appreciate fruitful discussions with Profs. S. Fujioka, M. Murakami and H. Azechi of ILE.

\end{document}